1

# Brand Intelligence Analytics

Fronzetti Colladon, A., Grippa, F.

# Brand Intelligence Analytics

**Andrea Fronzetti Colladon**[a, 1], **Francesca Grippa**[b]

[a] University of Perugia, Department of Engineering, Perugia, Italy, andrea.fronzetticolladon@unipg.it

[b] Northeastern University, Boston, MA, USA, f.grippa@northeastern.edu

*"In God we trust. All others must bring data."*

*W. Edwards Deming*

**Abstract –** Leveraging the power of big data represents an opportunity for brand managers to reveal patterns and trends in consumer perceptions, while monitoring positive or negative associations of the brand with desired topics. This chapter describes the functionalities of the SBS Brand Intelligence App (SBS BI), which has been designed to assess brand importance and provides brand analytics through the analysis of (big) textual data. To better describe the SBS BI's functionalities, we present a case study focused on the 2020 US Democratic Presidential Primaries. We downloaded 50,000 online articles from the Event Registry database, which contains both mainstream and blog news collected from around the world. These online news articles were transformed into networks of co-occurring words and analyzed by combining methods and tools from social network analysis and text mining.

## 1. Introduction: A Brand Intelligence Framework

In this paper, we describe a new dashboard and web app to assess brand image and importance through the analysis of textual data and using the composite indicator known as Semantic Brand Score (SBS) [1, 2]. The predictive power of the SBS and its three dimensions, i.e. prevalence, diversity and connectivity, has been demonstrated in various settings, including tourism management and political forecasting [1, 3].

Differently from traditional measures, the SBS has the benefit of not relying on surveys – which are usually subject to different biases e.g. [4–6]. The analysis is not constrained by small samples or by the fact that interviewees know that they are being observed. A set of texts could represent the expressions of an entire population, as for example all the news articles about Greta Thunberg. The SBS can be calculated on any source of text, including emails, tweets and posts on social media. The goal is to take the expressions of people (e.g. journalists, consumers, CEOs, politicians, citizens) from the places where they normally appear. This is aligned with previous research which proposed software and algorithms to mine the web, identify trends and measure the popularity of people and brands [7, 8].

[1] Corresponding author: Andrea Fronzetti Colladon email: andrea.fronzetticolladon@unipg.it

In this chapter, we describe how the SBS components can be fully translated into reports available to brand managers and digital marketing professionals. In order to demonstrate the benefits of the SBS BI app and describe some of the reports it generates, we apply the framework to the case of the 2020 US Democratic Presidential Primaries, by mining 50,000 online news articles and combining methods and tools of social network analysis and text mining.

In addition to the calculation of the SBS, the analysis we conduct is based on topic modeling, sentiment analysis and the study of word co-occurrences – which help reveal patterns and trends in consumer perceptions, identifying positive, neutral or negative associations of the brand with other topics [9]. The association between different concepts used in an online discourse to describe a brand can help marketing managers discern the perceived relationships among brands, as well as their positioning in the customers' mind.

## 1.1. The Semantic Brand Score

The Semantic Brand Score (SBS) [2] is a novel measure of brand importance, which is at the core of the analytics we describe in this chapter. It was designed to assess the importance of one or more brands, considering dynamic longitudinal trends using data from multiple online sources and different contexts. It is a measure suitable for the analysis of (big) textual data across cultural systems and languages. The SBS conceptualization was partially inspired by well-known brand equity models and by the constructs of brand image and brand awareness [10].

The concept of "brand" is very flexible and the SBS can be calculated for any word, or set of words, in a corpus. By 'brand' one could intend the name of a politician, or multiple keywords representing a concept (for example, the concept of "innovation" or a corporate core value). The measure was used to evaluate the transition dynamics that occur when a new brand replaces an old one [2], to evaluate the positioning of competitors, to forecast elections from the analysis of online news [1], and to predict trends of museums visitors based on tourists' discourse on social media [3].

The SBS has three dimensions: prevalence, diversity and connectivity. Prevalence measures the frequency of use of a brand name, i.e. the number of times a brand is directly mentioned – which can be considered a proxy of brand awareness and recall. Diversity measures the heterogeneity of the words associated with a brand, i.e. the richness of its lexical embedding. Connectivity represents a brand's ability to bridge connections between other words, which can represent concepts or discourse topics. The sum of these three indicators measures brand importance. The metric is fully described by the work of Fronzetti Colladon [2].

## 1.2. Textual Brand Image

When designing and evaluating brand-building programs, marketing managers should assess and measure brand image and brand equity by using other brands as benchmark, based on similarity of their positioning strategies. At the same time, many brand managers face the challenge to identify and visualize the correct measures of brand strength to complement financial measures with brand asset measures [11]. Building a strong brand image requires the adoption of a comprehensive measurement system able to validate brand-building initiatives and continuously monitor the impact on customer perception. An emerging method to assess and build individual brand image is the study of the words used to describe a brand.

Some scholars are adopting the theory of memetics to develop prediction tools to assess the spread of innovations [12] or to understand how concepts and brands are positioned in the minds of consumers. As noted by Marsden [13], how a brand is positioned in the associative networks of memory can be used to describe the meaning of that idea for customers. Techniques such as memetic analysis and use of brand mapping [14] allow marketing managers to assess how brands are positioned in the minds of consumers and whether these associations are positive, negative or neutral. These insights will support a better positioning of brands to fit with the consumers' mindset.

Measuring brand similarity is useful when selecting the most appropriate brand name, or to understand how a brand resemblance with another could impact brand loyalty and price sensitivity. Measuring brand similarity is also key when assessing how complete and comprehensive the information provided to customers is [15].

Traditional methods are usually based on surveys and use aggregated judgments made by potential customers [16]. Looking at the association between concepts describing a brand can help identify the perceived and psychological relationships among brands, their relative positioning and strategic differentiation.

Content analysis and topic modeling methods offer insights in the main topics discussed online, providing a set of keywords, and their connections [16, 17]. In this context, sentiment analysis of online data (e.g. news, reviews, blog entries) also comes to help and measures users' emotions and users' polarity towards a specific event, public figure or brand.

Recent studies have mined large-scale, consumer generated online data to understand consumers' top-of-mind associative network of products [18] by converting them into quantifiable perceptual associations and similarities between brands. Others have gone beyond the mere occurrence of terms in online data and assessed the proximity or similarity between terms using the frequency of their co-occurrence within a text [19]. Gloor and colleagues visualized social networks as Cybermaps and used metrics such as betweenness centrality and sentiment to evaluate the popularity of brands and famous people [7, 8].

## 2. The SBS BI Web App

The SBS Brand Intelligence App (SBS BI) has been developed to support the assessment of brand importance and the study of brand image and characteristics, through the analysis of (big) textual data[2]. This section describes the App's main components (version 4.5.10) and the analytical reports it generates.

As shown in Figure 1, the app has several menus, starting with the option to upload and analyze any text file that is available to the user. The app has also modules that allow the fetching of online news and tweets. The fetching modules use the Twitter API[3] and the Event Registry API [20] in order to collect data. A dedicated option gives users the opportunity to connect to the Telpress[4] platform, for the collection of news and the download of data which perfectly integrates with the SBS BI app. After uploading a csv file, the user is expected to set a number of parameters – such as the language and time intervals of the analysis, the word co-occurrence range, and the minimum co-occurrence threshold for network filtering (see [2] for more details). The last step consists of running the core module, which will calculate the SBS and the other measures described in Section 3.2.

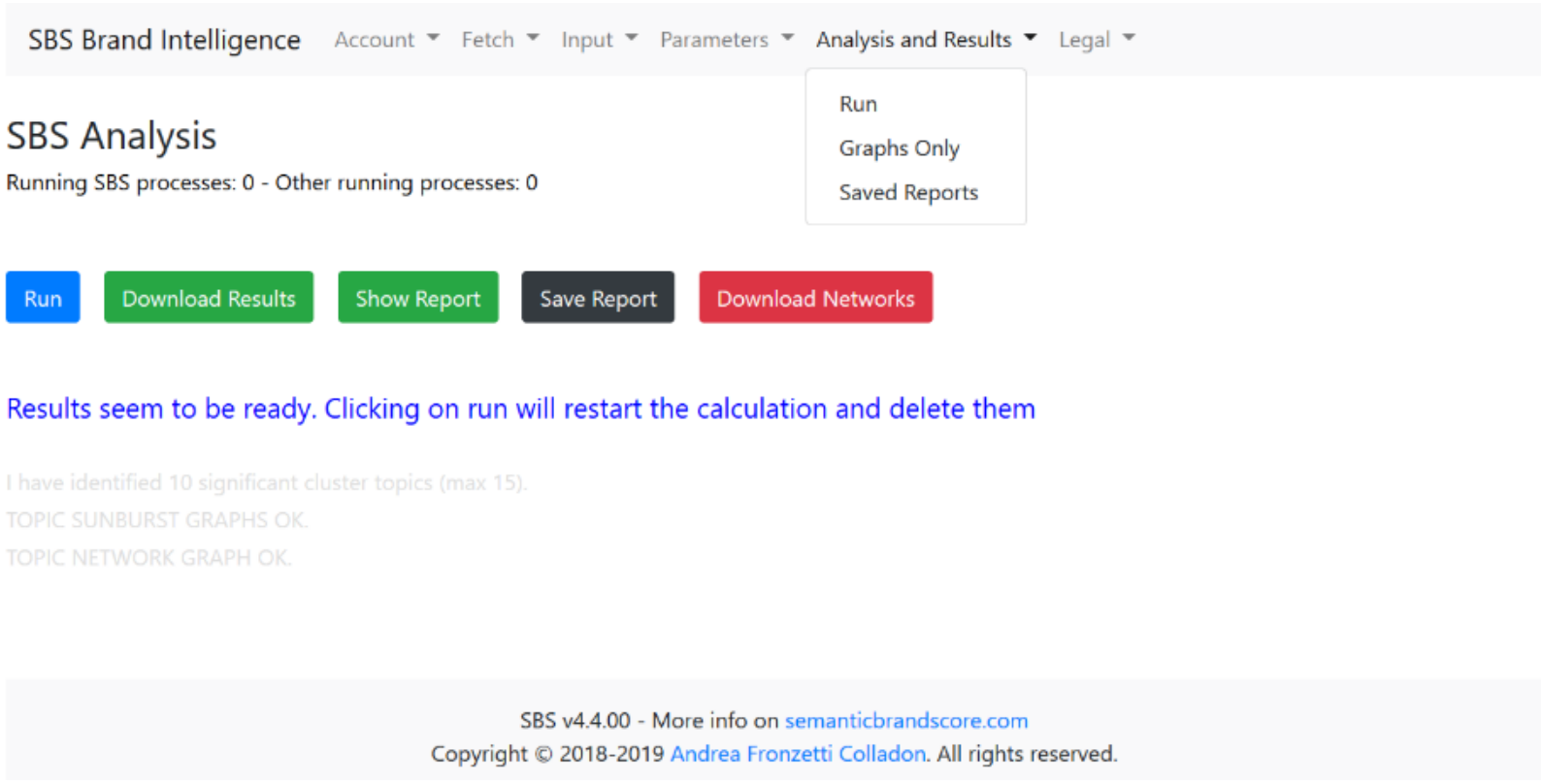

**Fig. 1** SBS Brand Intelligence App

---

[2] The SBS BI web app is distributed as Software as a Service, access can be requested for research purposes. Web address: https://bi.semanticbrandscore.com. Conceptualized and developed by Andrea Fronzetti Colladon (Copyright © 2018-2020).

[3] https://developer.twitter.com/en/docs/api-reference-index

[4] http://www.telpress.com/

## 2.1. Text Preprocessing

A preliminary step before the calculation of all metrics is the preprocessing of uploaded texts, starting with removing web addresses, punctuation, stop-words and special characters. Documents are subsequently tokenized and words are converted to lowercase. Word affixes are removed through the snowball stemmer included in the NLTK Python package [21].

After the preprocessing phase, documents are transformed into undirected networks, based on word co-occurrences. In these networks, nodes represent words and links among them are weighted based on co-occurrence frequencies. This serves to the calculation of the SBS. SBS BI gives users the option to download networks in the Pajek file format [22].

## 2.2. Calculation of the Semantic Brand Score

The SBS is the metric at the core of our analytics. Its dimensions of diversity and connectivity are calculated through the metrics of degree and weighted betweenness centrality [1, 2]. The traditional degree centrality metric can be adjusted [23] to value more the connections to low-degree nodes:

$$Diversity_i = \sum_{\substack{j=1 \\ j \neq i}}^{N} \log_{10} \frac{N-1}{g_j} \, I_{(w_{ij}>0)}$$

In the formula, $Diversity_i$ is the diversity of node $i$, $N$ is the total number of nodes in the network, $g_j$ is the degree of node $j$, and $I_{(w_{ij}>0)}$ is the indicator function which equals 1 if the edge connecting node $i$ to node $j$ exists, and 0 otherwise. We assume $w_{ij} = 0$ for unconnected nodes. The idea behind this adjustment is that associations of a brand are more distinctive if they occur with words having fewer connections. Several other variations of this metric are possible [23].

Prevalence is the count of word frequencies. Each measure is subsequently standardized, considering all the words in the network, by subtracting the mean to individual scores and dividing by the standard deviation. Standardized scores are added up to calculate the SBS. Other standardization techniques are also implemented by the app – such as min-max normalization or standardization obtained by subtracting the median and dividing by the interquartile range. Raw and standardized scores are provided as output.

Lastly, the SBS can also be calculated attributing different weights to different text documents. For example, the analyst might want to consider as more important an article published by The New York Times than one published by The Onion newspaper. Weights can be determined by the user and uploaded into the system. One possible approach for the determination of weights of online news is to refer to the Alexa[5] ranking of their sources. Other factors could also impact brand importance and should be considered, such as whether the article was published on the home page or not. According to this logic, if source A is 10 times more important than source B, prevalence of a brand mentioned by A will be 10 times higher than prevalence of a brand mentioned by B. Weights of network links are also determined considering source weights, and filtered accordingly.

Similarly, the analyst might want to limit the analysis to the initial part of online news, considering that most readers stop before reading 30% of webpages [24] and that a brand that appears in the title of an article is presumably more relevant than one only appearing at the end of its body. The SBS BI app offers the possibility of limiting the portion of text that will be analyzed.

## 2.3. Brand Intelligence Dashboard

Some of the most relevant information obtained from the analysis is summarized by a graphical dashboard. Some of the main graphs included in the dashboard are described in the following, whereas examples are provided in Section 4. All graphs are interactive and were created via the Plotly[6] library, excluding the topic network which has been generated using Cytoscape Js [25]. The app was mainly programmed using the Python language.

### *2.3.1. SBS Time Trends*

The SBS Time Trends interactive line graph shows the dynamic evolution of the semantic brand score for each brand over time. In a second tab, absolute values are replaced by proportional values with respect to competitors (see Figure 2).

### *2.3.2. Brand Positioning*

This is a scatter plot with the SBS on the vertical axis and brand sentiment on the horizontal axis (Figure 3). Combining information from these two measures we can have an idea of brand positioning, with the most important brands being located in the top right part of the graph (high importance and positive sentiment). Different approaches are possible for the calculation of sentiment. The app default for the English language is to use the VADER lexicon included in the NLTK library [26]. Sentiment varies between -

[5] https://www.alexa.com/siteinfo

[6] https://github.com/plotly/plotly.py

1 and +1, where -1 is negative and +1 is positive. It is important to notice that SBS BI calculates sentiment considering the sentences related to each brand, and not average scores of full documents. This is important for example to make a distinction in the case of a text where two brands are mentioned and the author is speaking in a positive way about a brand and negatively about the other. Punctuation is considered in the calculation of sentiment.

### 2.3.3. Average SBS Scores

The third graph offers a visualization of the average SBS scores obtained considering the full time of analysis (Figure 4). It is a stacked bar chart which shows the contribution of prevalence, diversity and connectivity to the final score. Each measure is rescaled in the interval [0, 100]. A more appropriate evaluation of the overall importance of each brand could also be obtained repeating the analysis on a single time interval, including all text documents available.

### 2.3.4. Most Common Words and Brand Associations

It can be interesting for the analyst to know the most frequent words used in a specific timeframe or overall, to discover concepts, people and events that were typically mentioned in a text corpus. SBS BI provides this information, through a dynamic sunburst graph. In addition, other charts show the top textual associations with the analyzed brands. Looking at the most frequent word co-occurrences, the user can understand the textual image of the brand, and its related message (Figure 5). In order to identify the main traits that distinguish a brand from competitors, the app also shows unique associations.

### 2.3.5. Brand Image Similarity

In this chart, the more similar is the textual image of two brands, the closer they appear (Figure 6). The user can get an overall view of the similarity of the words that co-occur with different brands. This can be seen as a proxy of the brand image of text authors, or could be used to assess similarity of communication strategies, if text are authored by companies. Cosine similarity is the metric used [27], together with multidimensional scaling [28], in order to plot the graph in two dimensions.

### *2.3.6. Target Words for Connectivity and SBS Improvement*

In addition to measuring the importance of a brand, it is also useful to understand what actions can be taken in order to improve this score. Prevalence increases if a brand is frequently mentioned. Accordingly, the press office of a company, or the campaign office of a political candidate, could work to obtain more media coverage. When diversity is low, its value can be increased by linking the brand name to heterogeneous themes and concepts. However, designing a strategy to improve connectivity, i.e. the brand 'brokerage power', is less easy. Brand managers need to find those words that, if used in future communication, could potentially make their brand more central in the discourse. However, they should also avoid favoring competitors and pay attention to keeping communication consistent with their brand strategy. In terms of graph theory, this is a maximum betweenness improvement problem [29], with additional constraints – such as the presence of forbidden nodes and opponents. Specific algorithms are implemented in the SBS BI app to solve this problem. The best set of words is shown by the target words graph (Figure 7) and it can be customized for each brand. These are the words that, if connected to a brand, have the highest potential to increase its connectivity.

### *2.3.7. Main Discourse Topics*

Topic modeling is a popular theme in text mining [17], with some of the most common approaches using Latent Dirichlet Allocation [30]. The goal is to automatically extract the main discourse topics from a set of documents and represent them through their most salient words. The SBS BI app reaches this goal using a different methodology, i.e. through the clustering of the full co-occurrence network. After the removal of isolates and negligible links, the Louvain algorithm [31] is used to determine the main network clusters (other approaches are also possible). Words that better represent each cluster are subsequently identified through the following formula:

$$IW_i^K = \sum_{\substack{j \in K \\ j \neq i}} w_{ij} \frac{\sum_{\substack{j \in K \\ j \neq i}} w_{ij}}{\sum_{\substack{j=1 \\ j \neq i}}^{N} w_{ij}} = \frac{\left(\sum_{\substack{j \in K \\ j \neq i}} w_{ij}\right)^2}{\sum_{\substack{j=1 \\ j \neq i}}^{N} w_{ij}}$$

where $IW_i^k$ is the importance of the word i belonging to the cluster *K*, *N* is the number of nodes in the network, and $w_{ij}$ is the weight of the arc connecting nodes $i$ and $j$ (other approaches are also possible). The idea is that the most representative words are those with many strong connections within the cluster and a low proportion of links to nodes outside the cluster – similar to the logic of modularity functions [32]. Figure 8 shows the topic modeling graph. This graph also helps identify which topic is closest to each brand (red nodes) and the strength of connection between the topics. The app also calculates the importance of each topic and the weight of its connections to the different brands.

Lastly, some other charts are produced by the app, such as the time trend of the number of unique brand associations. The App allows the analyst to generate new, customized reports using the results files, which can be downloaded at the end of the analysis.

## 3. Case Study: 2020 US Democratic Primaries

The field of 2020 Democratic presidential candidates has been defined by many commentators as the largest Democratic primaries field in modern history, since it involved more than two dozen candidates and included six female candidates. As of November 24, 2019, a total of 18 candidates were seeking the Democratic presidential nomination in 2020.

On November 26, 2019 we downloaded 50,000 articles from the Event Registry database [20] – which contains both mainstream and blog news collected from around the world. We selected the most recent articles which were related to the 2020 Presidential Race and the Democratic Primaries. Articles were published in the USA in the period November 10-25, 2019. Using the SBS BI App, we generated reports for the top four candidates that had a vote share higher than 5% in the last available national polling average [33]. These candidates were Joseph R. (Joe) Biden Jr., Elizabeth Warren, Bernie Sanders and Pete Buttigieg.

Figure 2 illustrates the Time Trends interactive graph for the selected candidates, with fluctuating dynamic positions over time. We notice that Biden’s positioning is constantly higher than the others, which indicates a higher frequency with which the Biden name appears in the online news, but also a higher diversity and connectivity. This can be explained by the events associated with Biden’s son in Ukraine, the subsequent impeachment process for President Trump, in addition to the internal discussion with the rest of the primaries candidates. Conversely, the SBS trends for the others are lower, though with a higher fluctuation. In particular, SBS trends for Buttigieg and Sanders are more intertwined, which might indicate that online news report stories about them that are highly associated.

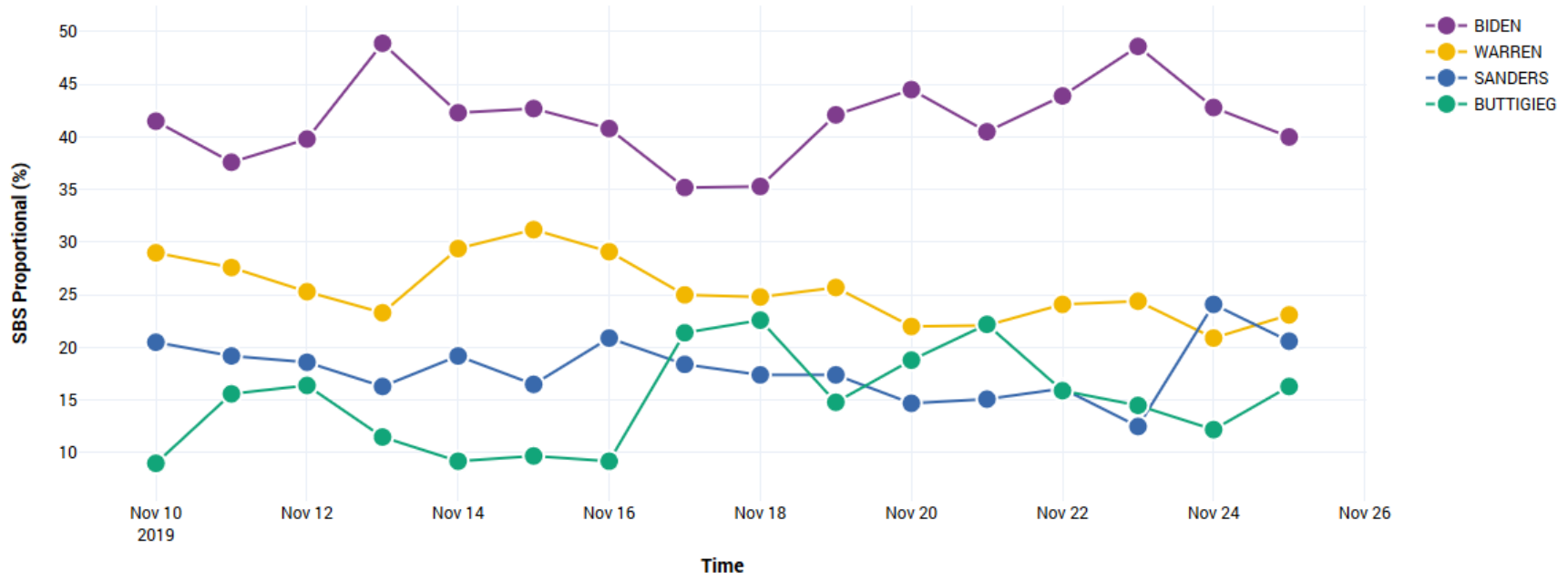

**Fig. 2** SBS proportional time trends

The scatter plot in Figure 3 combines information from the SBS values (y axis) and the brand sentiment values (x axis). The online discourse around Biden is the most different in terms of variety of news reported about him. Warren is the second most reported candidate with Buttigieg and Sanders immediately after. In terms of sentiment analysis, the online news present the front-runner (Biden) in a more neutral way, perhaps due to the Trump-Ukraine scandal and his son's involvement overseas. While Biden is associated with a more diverse set of topics (both positive and negative), the other three candidates are associated with more positive words.

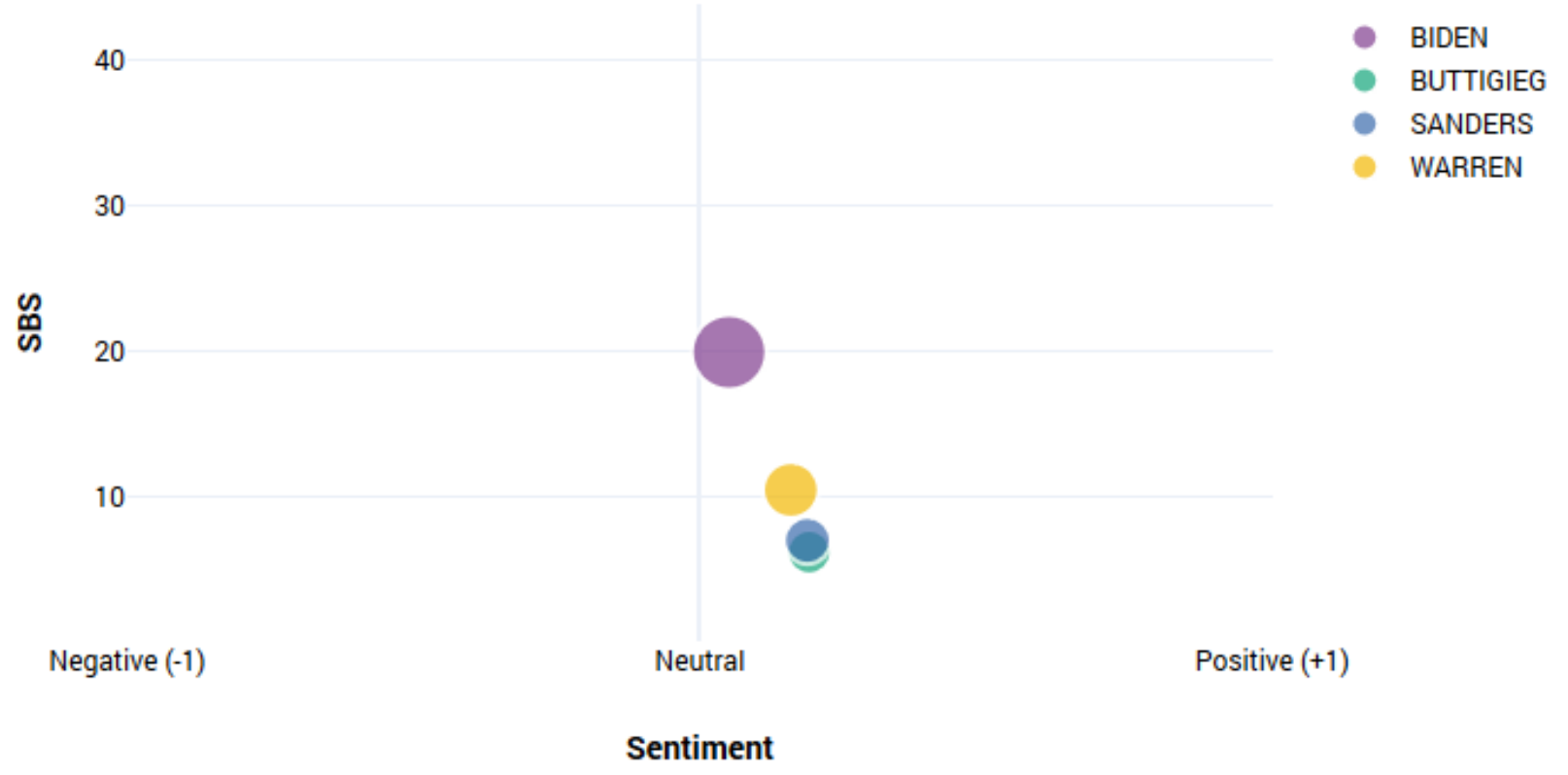

**Fig. 3** Brand positioning

Figure 4 shows the stacked bar chart with the average SBS scores during the entire time interval (November 10-25). We notice that Sanders has the lowest relative value of connectivity, which means that the brand "Sanders" serves fewer times as an indirect link between all the other pairs of words in the co-occurrence network. In other terms, the Sanders brand does not support an indirect connection between political concepts that are not directly co-occurring. This could indicate that Sanders was reported on via online media as talking about a specific set of agenda points, without much connection to other clusters of concepts. On the contrary, the Biden brand has the highest indirect link between all the other pairs of words, as it appears on the media as connecting clusters of concepts that are not directly connected to each other.

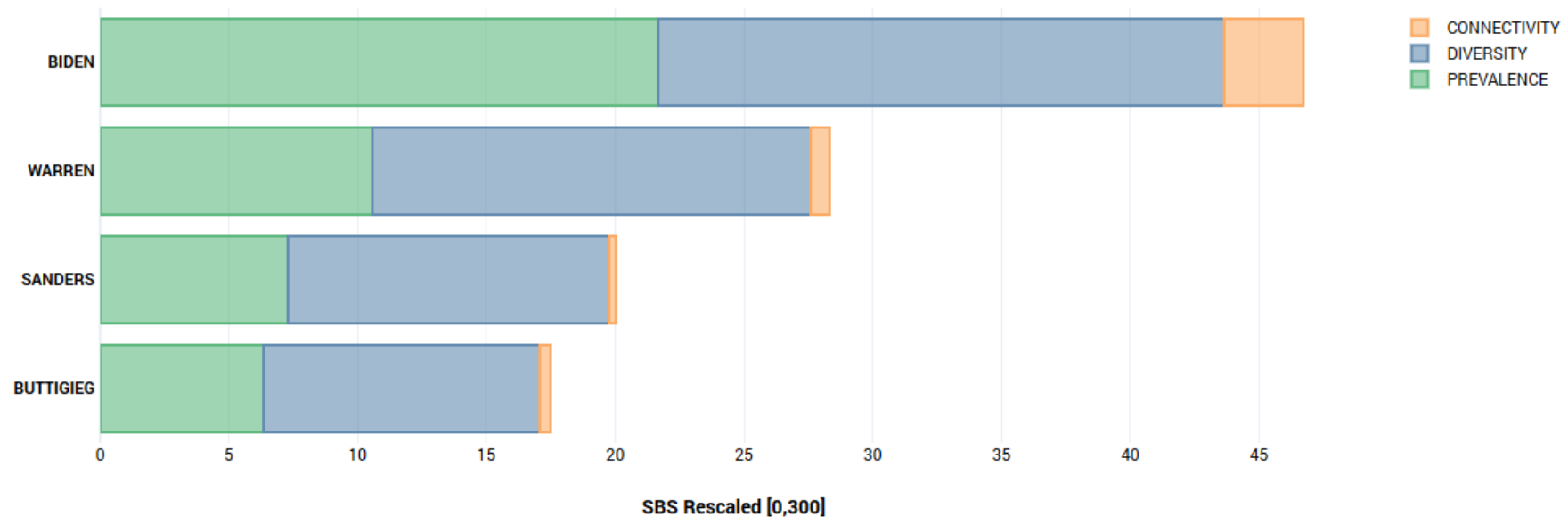

**Fig. 4** Average SBS scores

The textual brand associations, which are illustrated in Figure 5, present some interesting insights for the selected candidates. If we zoom into the discussion around Joe Biden we notice some of the most frequently used words in the specific timeframe: "Burisma", "Hunter", "investigation", and "son" were all more frequently mentioned with the candidate "Biden" in the online news. It is not a surprise that these were the top textual associations with the "Biden-brand", since Burisma is the holding company for a group of energy exploration and production companies based in Ukraine where Biden's son, Hunter, worked and was a board member. The topics associated with the other candidates were more diverse and refer directly to the specific agenda points that the candidate would bring to the table in a future presidential race. For example, the textual brand association of Elizabeth Warren highlights largely discussed points such as "Medicare, wealth, and billionaire", which are indeed the key differentiators of the Massachusetts senator. The textual association for Sanders was somewhat similar to Warren's, as Sanders co-occurred with concepts such as "progressive and Medicare".

If we look at the unique associations of words to each brand/candidate, it is interesting how Buttigieg has a strong association with "combat, nonwhite, qualified and gay". This is not surprising since Buttigieg – former Mayor of South Bend, Indiana, and a veteran of the War in Afghanistan – would be the first openly gay president, if elected, and has received strong support from the "non-white" part of the Democratic base. Former President Barack Obama once called him the "future of the Democratic Party".

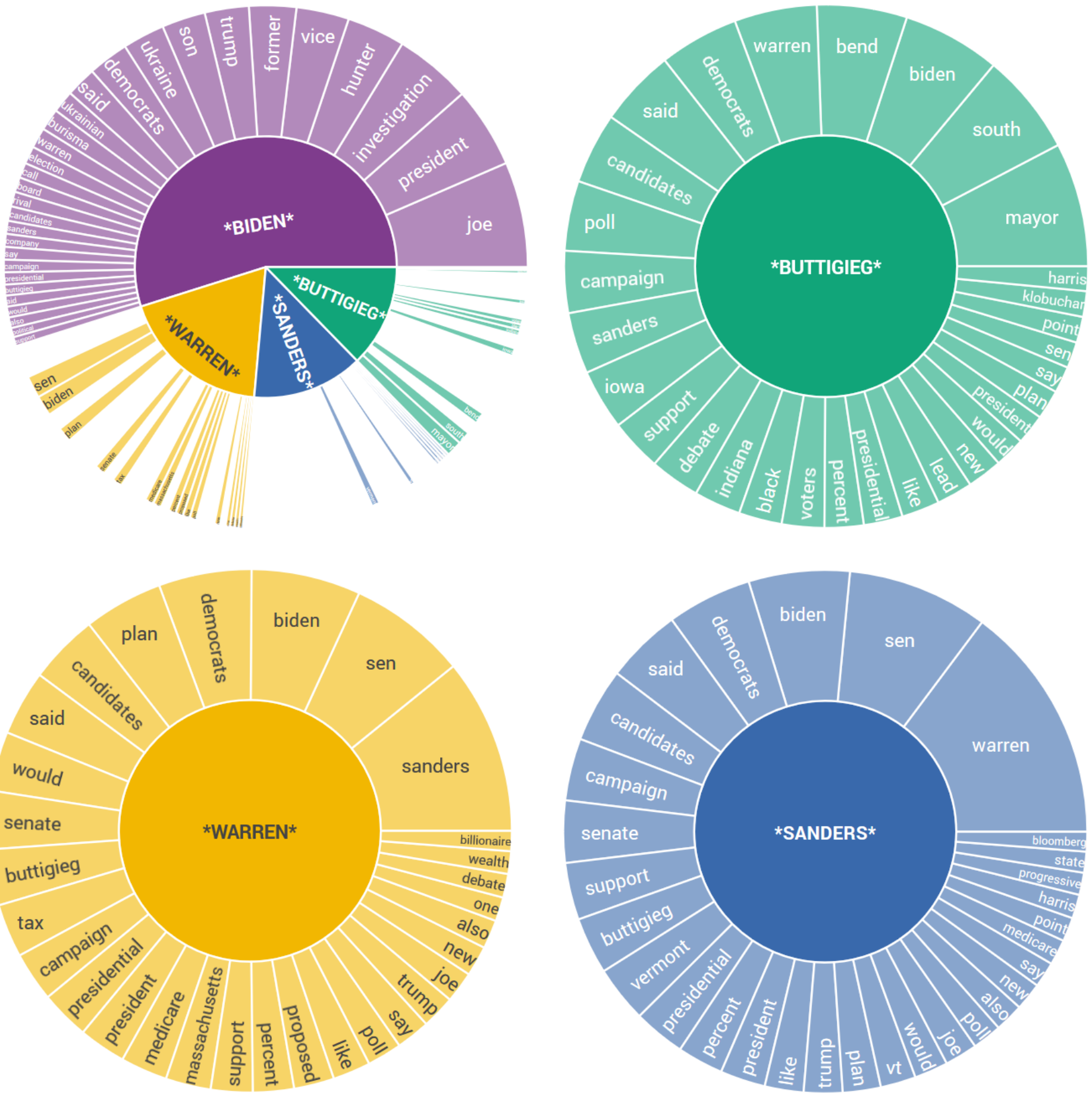

**Fig. 5** Textual brand associations

The report offers also some insights in terms of image similarity. The graph in Figure 6 indicates that in terms of image similarity Biden differentiates himself in a more significant way, likely because of the candidate's association with the Ukrainian investigation involving the current President, Donald Trump. Warren, Buttigieg and Sanders tend to be reported more often as sharing a similar political discourse. This is an important insight for their political campaign since voters might not be able to distinguish one candidate from the other unless they become more specific with their positions.

Sanders, Buttigieg and Warrens' positions on mental health and health policy show how their images are similar to each other and far from Biden. While the candidates have focused their dominant themes on universal health care, climate change, and reproductive rights, Vice President Joe Biden has been slower to embrace marijuana law reform and the legalization of cannabis for medical purposes [34], which could help veterans avoid or alleviate substance abuse disorder.

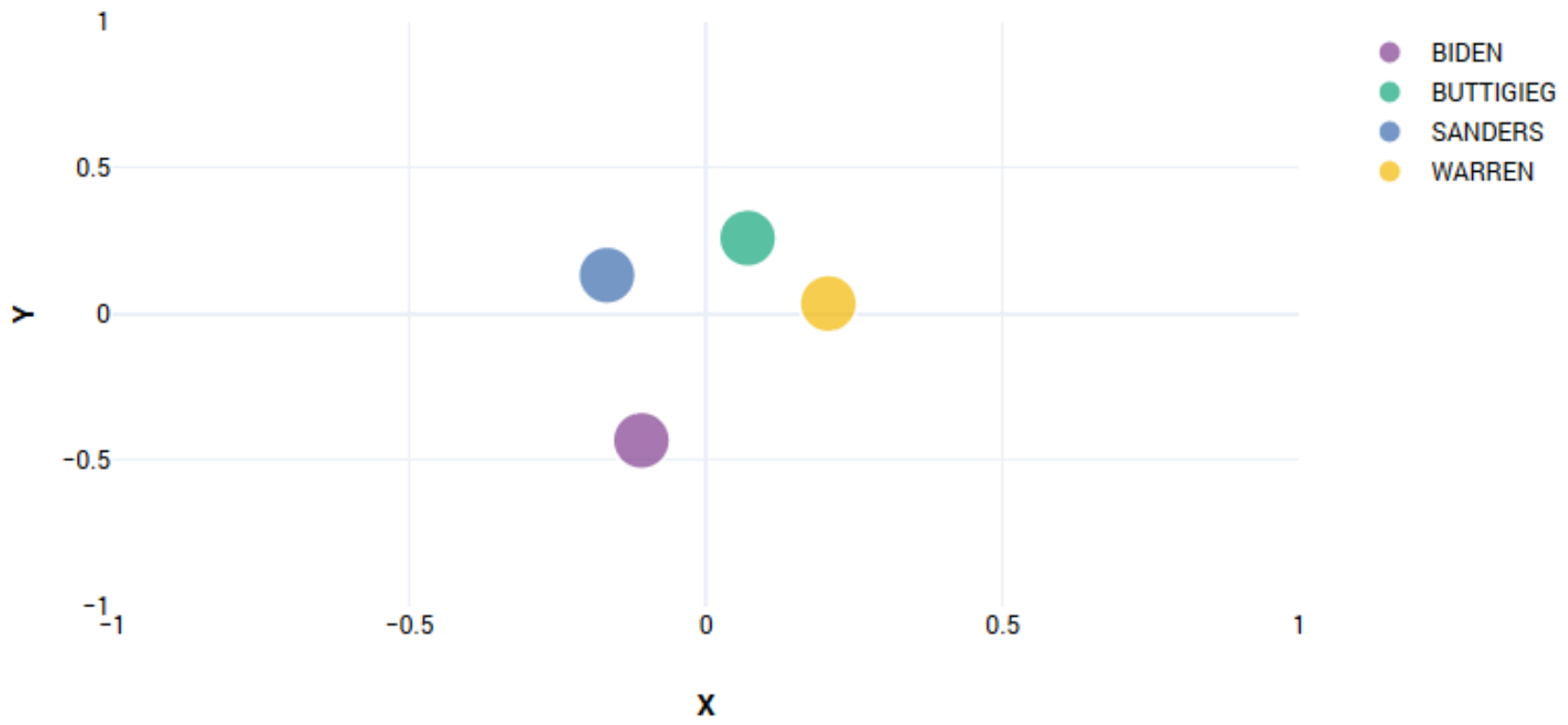

**Fig. 6** Image similarity

The target words graph in Figure 7 illustrates how the online discourse of these candidates is predominantly focused on the current President, Mr Trump. The online press is frequently reporting the direct statements of the politicians (see the highly reported word “said”), as well as topics such as “impeachment, Ukraine, State, House (of Representatives)”. These concepts act as catalyst of connectivity between the candidates and the rest of the political discourse.

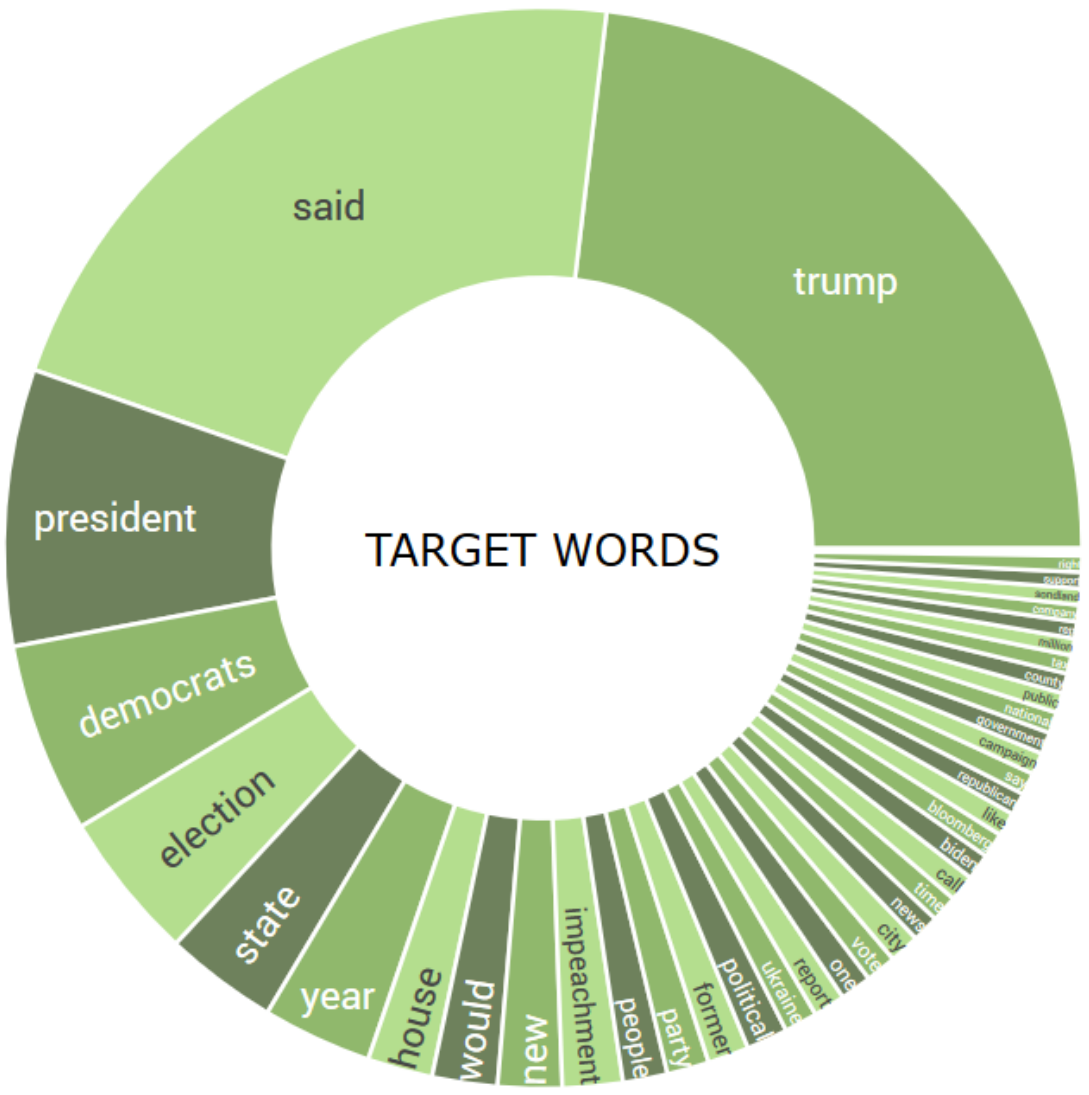

**Fig 7** Target Words

Figure 8 shows the topic modeling graph. The network of keywords indicates what the main discourse topics are and which is closer to each brand. The brighter red nodes are the brand/candidates. Very interestingly, Biden is mainly embedded in a cloud of words (T1) that is separate from the clouds where Warren, Buttigieg and Sanders are reported (T2). The width of the link connecting T1 and T2 is big enough to suggest the all four candidates are associated to and talking about a sub-set of shared concepts. However, the nature of the words associated with Joe Biden shows a strong dissimilarity from the ones associated with the other candidates. Biden is reported in articles associated with "testify, impeachment, hearings, ambassador", which clearly refer to the Trump-Ukraine scandal. The cluster in which the other candidates are mainly embedded into is characterized by words associated with the primaries vote and election.

Another interesting insights from the topic modeling network is the extremely weak connection between the four candidates' clusters (T1 and T2) and the cluster T6, reporting topics such as "Israel strikes in Gaza, rockets, police action in Hong Kong, political unrest in Bolivia". This seems to suggest that all four candidates are currently focusing on national policies rather than foreign policy issues. We would expect that, at a later stage of the primaries process, the emerging front runners will be asked their opinion on foreign policy issues, which are important if they plan to become the next President of the United States.

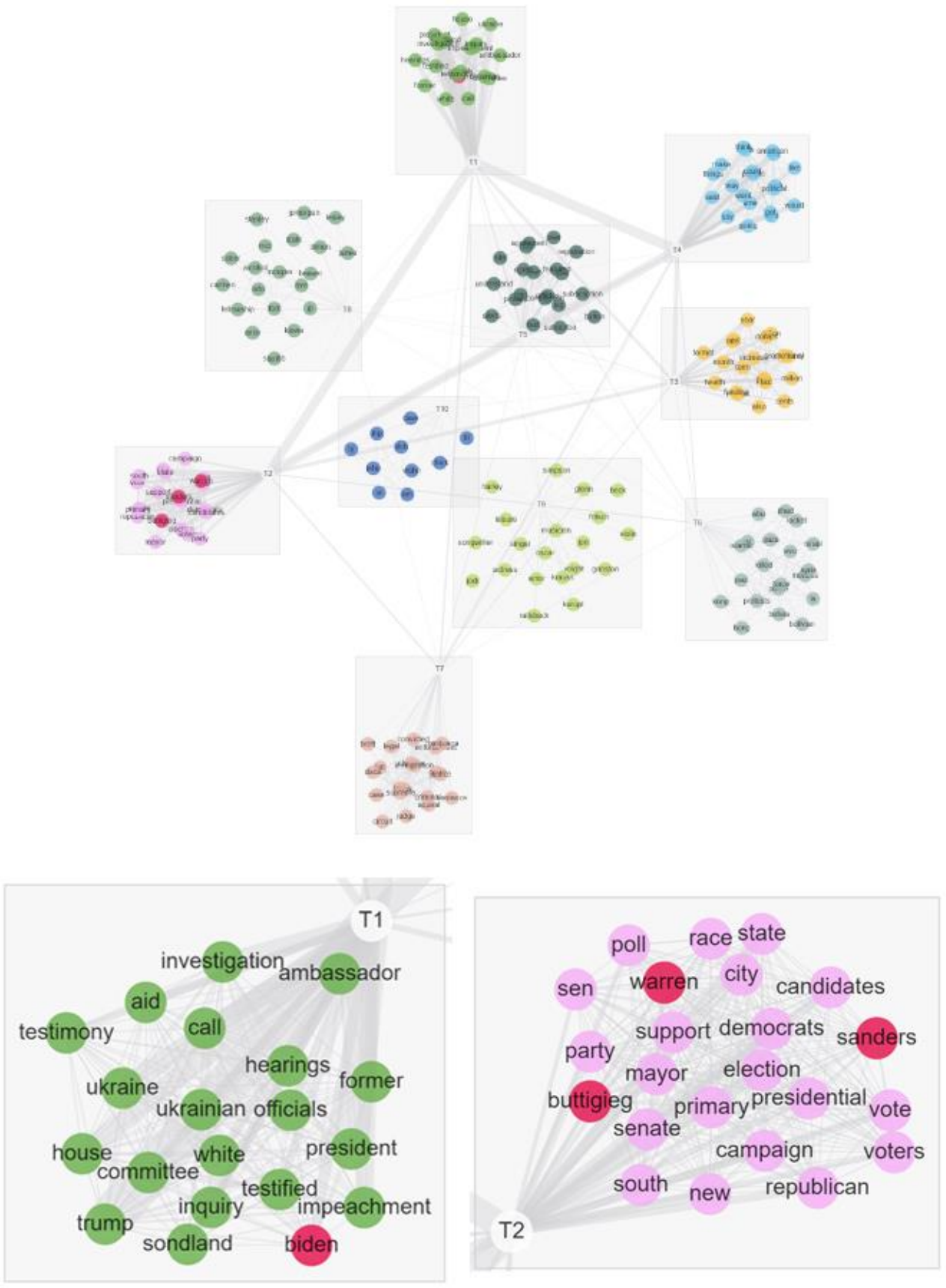

**Fig 8**. Main discourse topics.

Table 1 offers a comparison of brand importance in online news and poll results. The first column represents the percentages for each candidate based on the most recent poll. In the second column we reported the same values calculated as if these four candidates were the only ones to run. The third column reports the values of the predictions using SBS dimensions based on the online news sources [1]. The last column is the most interesting as it illustrates the difference between the values of the adjusted polls and the proportional SBS. This seems to indicate that the online discourse around these four candidates is translating into different proportional and relative impacts on their polls. While Pete Buttigieg is driving relatively fewer voters based on the polls, the media is reporting him as relatively more important and connected to a variety of topics.

| Candidate | Average Last Polls | Adjusted Polls | Proportional SBS | Difference (Prop. SBS - Adj. Polls) |
|---|---|---|---|---|
| Biden Jr. | 27% | 36.0% | 40.0% | 4.0% |
| Warren | 22% | 29.3% | 23.1% | -6.2% |
| Sanders | 18% | 24.0% | 20.6% | -3.4% |
| Buttigieg | 8% | 10.7% | 16.3% | 5.6% |

**Table 1:** Comparing brand importance in online news and poll results

## 4. Discussion and Conclusions

In this chapter, we have presented new methods to measure and assess the importance and relative positioning of brands. To explain the functionalities and reports available via the SBS BI app, we have discussed how the four front runners for the US Democratic 2020 primaries are positioned in the online news.

The SBS BI App represents an innovative tool to measure brand importance and brand positioning, combining the components of the SBS indicator (prevalence, diversity and connectivity) and relying on methods and tools of text mining, sentiment analysis and social network analysis.

Overall, the application of the SBS BI to a limited time period of the US democratic primaries indicates that Joe Biden is the one with the richer textual embedding, spanning boundaries of political discourse. The method we describe in this paper has the potential to complement traditional polls, by providing a comprehensive analysis of what people (news reporters, but also commentators, voters etc.) say about the candidates online. Our method is based on the automatic mining of big (textual) data, which could help counteract the so called “pollster fatigue”, where voters start to avoid answering the calls of pollsters, impacting the representativeness of the sample.

The SBS BI app – not limited to the analysis of political news – is in continuous development and we plan to add more functionalities in the near future. For example, we plan to improve the algorithm used for the identification of target words, to enrich the set of recommendations the app can provide to increase brand importance. Topic modeling through the clustering of co-occurrence networks still has open research questions, as well as the identification of the most salient words for each topic.